\documentclass[12pt]{article}
\usepackage{makeidx}
\usepackage{epsfig}
\usepackage{graphicx}
\usepackage{bm}
\usepackage{amsfonts}
\usepackage{latexsym}
\def\RR{\Bbb R}
\def\CC{\Bbb C}

\begin{document}
\begin{titlepage}
\bigskip
\rightline{} 

 \bigskip\bigskip\bigskip\bigskip
 \centerline{\Large \bf {Topology change in quantum gravity\footnote{
 To appear in the proceedings of Stephen Hawking's $60^{th}$
 birthday conference, Cambridge University, Jan. 2002}}}
 \bigskip\bigskip
 \bigskip\bigskip
 \centerline{\large Fay Dowker}
 \bigskip\bigskip
 \centerline{\em Department of Physics, Queen Mary, University of London,} 
\centerline{\em Mile End Road, London E1 4NS, United Kingdom}
 \bigskip\bigskip

A particular approach to topology change in
quantum gravity is reviewed, showing that several aspects of Stephen's
work are intertwined with it in an essential way. Speculations
are made on  possible implications for the causal set 
approach to quantum gravity.
 
 \end{titlepage}

\section{Introduction}

The challenge facing anyone giving a talk in celebration of
Stephen Hawking's contributions to physics is to convey something
of the epic breadth of his work in a very short time.  
The subject of topology change in quantum gravity provides an 
opportunity to do just that.
Topology change is not only a subject  
to which Stephen has directly made 
many seminal contributions 
but also one which, looked at from a particular point of view,
weaves together several  major themes of his work over the years. 
This is the point of view I will 
describe.  

The framework for topology change I will set out exists in 
what might be called a ``top down''  approach to quantum gravity. 
By this I mean that we take what we know -- general relativity, 
continuum spacetime, quantum field theory and so forth -- and
try to put them together as best we can, preserving as much structure
as possible. 
Most workers in the field believe that this 
approach will not, ultimately,
be adequate for quantum gravity and that we will need a 
new ``bottom up''
theory for which we postulate new fundamental structures and principles 
and in which continuum spacetime is an emergent phenomenon.   
Of course it may turn out that the rules we now choose to apply to 
our top down calculations will be shown to be wrong when we 
have the underlying theory of quantum gravity, there is that risk, 
but as a strategy it 
seems unavoidable. How could we hope to deduce the new principles 
completely unguided?  We proceed in the hope that our top down calculations 
are giving us clues in our search for the correct bottom up theory. 
And contrariwise, our beliefs about the eventual form of the bottom up 
theory will inform the choices we must make in those top down calculations. 
There's a rich and complicated 
process of cross-influence involved in the whole 
endeavour.  

In this spirit, we can hope that topology change will be a 
particularly fruitful area to study, since it is generally 
believed that it does not occur in classical general relativity and 
so would be a genuinely quantum  phenomenon -- one in which the 
underlying theory must leave its indelible mark 
in the realm of the continuum. 

I have not attempted to give a comprehensive set of 
references but have tended to cite papers where more details can be
found on some of the main arguments. 

\section{A top down framework for topology change}

What we will mean by a topology change is a
spacetime based on an n-dimensional manifold, $M$, with 
an initial spacelike $n-1$ dimensional hypersurface, 
$\Sigma_0$, and a final spacelike hypersurface, $\Sigma_1$,
not diffeomorphic to $\Sigma_0$. 
 For simplicity in what 
follows we take $\Sigma_0$ and $\Sigma_1$ to be closed
and $M$ compact so that the boundary of $M$ is the disjoint union of
$\Sigma_0$ and $\Sigma_1$. This restriction   
means effectively that we're studying topology changes that 
are localized. For example, a topology change from $\RR^3$ 
to $\RR^3$ with a handle -- an $S^2 \times S^1$ -- attached 
can be reduced to the compact case because infinity in 
both cases is topologically the same. On the other hand,  
a topology change from $\RR^3 \times C$   
to $\RR^3 \times C'$ where $C$ and $C'$ are  two non-diffeomorphic 
Calabi-Yau's, say, will not be able to be accommodated in the 
current scheme because it involves a change in the topology of infinity.

Following Stephen \cite{Hawking:1978jz, Hawking:1978zw}, 
I prefer the Sum Over Histories 
(SOH) approach to quantum gravity which can be summarized in the 
following formula for the transition amplitude between 
the Riemannian metric
$h_0$ on $(n-1)$-manifold $\Sigma_0$ and the Riemannian metric
$h_1$ on  $(n-1)$-manifold $\Sigma_1$:
\begin{equation} 
\label{eqn:soh}
\langle h_1 \Sigma_1 \vert h_0 \Sigma_0 \rangle
= \sum_{M} \int_g [dg] \omega(g)\ .
\end{equation}
The sum is over all $n$-manifolds, $M$, called cobordisms --
whose boundary is the disjoint union of $\Sigma_0$ and $\Sigma_1$ 
and the functional integral is 
over all metrics on $M$ which restrict to $h_0$ on $\Sigma_0$ and
$h_1$ on $\Sigma_1$. Each metric contributes a weight, $\omega(g)$, to 
the amplitude and we'll hedge our bets for now on the type of metrics
in the integral and hence the precise form of the weight. 

It's clear from this that the SOH framework lends itself to the study of 
topology change as it readily accommodates the inclusion of 
topology changing manifolds in the sum. Despite the
fact that 
we may not be able to turn (\ref{eqn:soh}) into a mathematically
well-defined object within the top down approach,  
if even the basic form of this transition amplitude 
is correct then we can already draw some 
conclusions. We can say that a topology change 
from $\Sigma_0$ to $\Sigma_1$ can only occur if there is at least one
manifold which interpolates between them, in other words if they are 
cobordant. This does not place any restriction on topology change in 
3+1 spacetime dimensions since all closed three-manifolds are cobordant,
but it does in all higher dimensions: not all closed four-manifolds  
are cobordant, for example. 
We can also say that even if cobordisms exist, there must also exist 
appropriate metrics on at least one cobordism and so we come to the 
question of what the metrics should be. 
There are many possibilities and  just three are listed here. 

\begin{enumerate}

\item Euclidean ({\it i.e.} positive definite signature) metrics. This choice is of course closely associated with Stephen and the whole programme of Euclidean 
quantum gravity \cite{Gibbons:1994cg}. It is to this tradition 
and to Stephen's influence that I attribute 
my enduring belief that topology change does occur in quantum gravity. Indeed,
Euclidean (equivalently, Riemannian) metrics exist
on any cobordism and it would seem perverse to exclude different topologies
from the SOH.  

\item Lorentzian ({\it i.e.} $(-,+,+,\dots +)$ signature) metrics. 
With this choice we are forced, by a theorem of Geroch \cite{Geroch:1967}, 
to contemplate closed timelike curves (CTC's or time machines). 
Geroch proved that if a Lorentzian metric exists 
on a topology changing cobordism then it must contain CTC's 
or be time non-orientable. 
Stephen has been at the forefront of the study of these causal pathologies, 
formulating his famous Chronology Protection Conjecture \cite{Hawking:1992nk}
(see also Matt Visser's 
contribution to this volume). Stephen and Gary Gibbons also proved that 
requiring an $SL(2,\CC)$ spin structure for fundamental fermi
fields on a Lorentzian cobordism 
produces a further restriction on allowed topology changing transitions
\cite{Gibbons:1992tp, Gibbons:1992he}.  

\item Causal metrics. By this I mean metrics which give rise to a well-defined
``partial order'' on the set of spacetime events. 
A partial order is a binary relation, $\prec$, on a set $P$,
with the properties:
\begin{itemize}
\item (i) transitivity:
  $(\forall x,y,z\in P)(x\prec y\prec z\Rightarrow x\prec z)$
\item (ii)  irreflexivity:
  $(\forall x\in P)(x\not\prec x)$\ . 
\end{itemize}
A Lorentzian metric provides a partial order 
via the identification $x \prec y \Leftrightarrow x \in J^-(y)$, where the 
latter condition means that there's a future directed curve 
from $x$ to $y$ whose tangent 
vector is nowhere spacelike (a ``causal curve''), 
so long as the metric contains no closed causal  curves. 
The information contained
in the order $\prec$ is called the ``causal structure'' of the spacetime. 
By Geroch's theorem, we know there are 
no Lorentzian metrics on a topology changing cobordism 
that give rise to a well-defined causal structure. But, there 
are metrics on any cobordism 
which are Lorentzian {\it almost} everywhere which do \cite{Sorkin:1989ea}. 
These metrics avoid Geroch's theorem by being degenerate at a finite 
number of points 
but the causal structure at the degenerate points is
nevertheless meaningful. Describing these metrics will be the job of the 
next section. 
\end{enumerate}

We will plump for choice 3, causal metrics, in what follows. There are many 
reasons to do so but the one I would
highlight is that it is the choice which fits in with a particular
proposal for the underlying theory, namely causal set theory, in 
which it is the causal structure of spacetime, over all its other 
properties,  that is primary and will persist at the 
fundamental level. We will return to causal sets later.  
This choice, of causal spacetimes in the SOH, 
cannot be deduced logically from any top down considerations. 
It is a choice informed by a vision of what kind of theory
quantum gravity will be when we have it. 

\section{Morse metrics and elementary topology changes}

Morse theory gives us a way of breaking a cobordism into 
a sequence of elementary topology changes \cite{Yodzis:1973, Alty:1995xs, Sorkin:1989ea}.
On any cobordism $M$ there exists a Morse function, 
 $f:M \rightarrow [0,1]$,
with $f|_{\Sigma_0} = 0 $, 
$f|_{\Sigma_1} =1$ such that $f$ possesses a set of critical
points $\{p_k\}$ where $ \partial_a f|_{p_k}=0$ and the Hessian,
$\partial_a\partial_bf|_{p_k}$, is invertible. These critical
points, or Morse points,
of $f$ are isolated and, because $M$ is compact, there are
finitely many of them.
The index, $\lambda_k$, of each Morse point, $p_k$ is the 
number of negative eigenvalues of the Hessian at $p_k$. It is the
number of maxima in the generalized saddle point at $p_k$ if 
$f$ is interpreted as a height function. For spacetime dimension 
$n$, there are $n+1$ possible values for the indices, 
$(0,1,\dots n)$. A cobordism with a single Morse point is 
called an elementary cobordism.

\begin{figure}[thb]
\centerline{\epsfbox{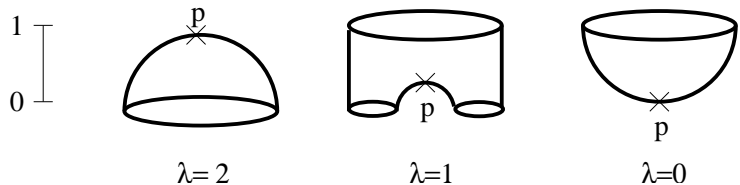}}
\caption{
Three elementary cobordisms for $n=2$ and $\lambda = 2,1,0$:
the yarmulke,  trousers and time-reverse of the yarmulke. 
}
\label{fig:elementary}
\end{figure}

Three elementary cobordisms for $n=2$ are shown in 
figure \ref{fig:elementary}.
They are the $\lambda=2$ yarmulke in which a circle is destroyed, 
the $\lambda=1$ trousers in which two circles 
join to form a single circle and the $\lambda=0$ time reverse of
the yarmulke in which a circle is created from 
nothing. (The upside-down trousers is in fact also a 
$\lambda =1$ elementary cobordism: locally the Morse point 
looks the same as the regular trousers with one maximum and one 
minimum.) 
For higher spacetime dimensions, $n$, the generalizations of these are 
easy to visualize: the index $n$ yarmulke (or its time reverse
of index $0$) is half an $n$-sphere, the index 1 trousers (or its time
reverse of index $n-1$) is an $n$-sphere with three balls deleted creating
three $S^{n-1}$ boundaries. For $n>3$  qualitatively different types of 
Morse point exist with at least two maxima and two minima, {\it i.e.} 
$\lambda \ne 0,1,n-1,n$. These are impossible to draw but we will 
encounter some examples in the next section. 

Using a Morse function, $f$, on $M$ we can construct ``Morse metrics'' 
which  are Lorentzian everywhere except at the Morse points where
they are zero. The precise form is not important here, but 
roughly the Morse function is used as a time function as you'd
expect. These Morse
metrics are our candidates for inclusion in the SOH for quantum 
gravity. 

Now, 
there is important counterevidence  to the claim that topology change
occurs in quantum gravity. This is work which shows that the expectation
value of the energy-momentum tensor of 
a massless scalar field propagating on a $1+1$ Morse trousers 
is singular along the future light cone of the Morse point \cite{Anderson:1986ww,
Manogue:1988}.
In addition, one can look instead at the in-out matrix element of the 
energy momentum tensor and one finds a singularity along both the 
future and past light cones of the Morse point (calculation 
described in \cite{Sorkin:1989ea}). This last result in particular,
if it can be extended to all Morse metrics on the $1+1$ trousers,
can be taken to suggest that in the full SOH expression for 
the transition amplitude, integrating out over the scalar field 
first will leave an expression for the effective action 
for $g$ that is infinitely sensitive to fluctuations in 
$g$. Thus, destructive interference between nearby metrics will
suppress the contribution 
of any metric on the trousers. 
To be sure, this is a heuristic argument that 
would need to be strengthened but suppose it is valid. Would
this mean, as DeWitt has argued, that all topology change is suppressed?
The answer is not necessarily, 
especially if the following two conjectures hold. 

The first conjecture is based on the idea that 
it is a certain property,
called ``causal discontinuity'', of the 
causal structure of the $1+1$ trousers that is the origin of the 
bad behaviour of quantum fields on it. So the conjecture (Sorkin) is that 
quantum fields will be singular on causally discontinuous spacetimes
but well-behaved on causally continuous spacetimes. 
The second conjecture (Borde and Sorkin) is that only Morse metrics
containing index $1$ or $n-1$ points (trousers type) are
causally discontinuous. 

\begin{figure}[thb]
\centerline{\epsfbox{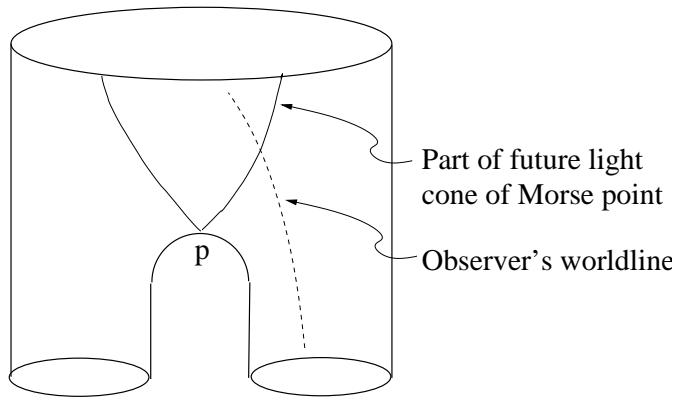}}
\caption{The $1+1$ trousers with part of the future light cone of 
the Morse point, $p$. The other part goes up the back. The 
past light cone of the Morse point also has two parts, one down
each leg.  
}
\label{fig:trousers}
\end{figure}

So what is causal discontinuity? When I was first learning about these things
I was excited to discover that Stephen himself invented the 
concept in work with R.K. Sachs \cite{Hawking:1974}. That paper is 
a piece of hard mathematical physics but 
there is a physically intuitive way of understanding 
the concept. Roughly, a spacetime is causally discontinuous if the 
causal past, or future, of a point changes discontinuously as the
point is moved continuously in spacetime. We can see 
from figure 
\ref{fig:trousers}
that it is 
very plausible that the 
$1+1$ trousers is causally discontinuous: an observer down in one 
of the legs will have a causal past that is contained only in that 
leg, but as the observer moves up into the waist region, as they pass
the future light cone of the Morse point, their causal past will 
suddenly get bigger and include a whole new region contained in the
other leg. 
Interestingly, Stephen and R.K. Sachs conclude
their paper by saying, ``[T]here is some reason, but no
fully convincing argument, for regarding causal continuity as a basic
macrophysical property.'' Which view ties in nicely with the first of
the two conjectures.  

\section{Good and bad topology change}

If we assume the two conjectures of the previous section 
hold, and that the argument about the consequent suppression of 
causally discontinuous metrics in the SOH is valid, the implication is 
that cobordisms which 
admit only Morse metrics containing index $1$ and/or $n-1$ points 
will be suppressed in the sum over manifolds. 
We can use this to draw conclusions about many 
interesting topology changing processes in quantum gravity. They 
divide into ``good'' ones which can occur and ``bad'' ones which 
do not.  
The results in this section and the next are taken from a series of papers 
on topology change \cite{Dowker:1998kc,Dowker:1998hj,
Borde:1999md,Dowker:1999wu}.

Good processes include the pair production of black holes of
different sorts in a variety 
of scenarios worked on by many people including Stephen (see the
contribution by Simon Ross in this volume). For example, in 
4 spacetime dimensions, the manifold  of the 
instanton that is used to calculate the 
non-extremal black hole pair production rate \cite{Garfinkle:1991eq}
admits a Morse function with a single index 2 Morse point. 
The pair production of Kaluza-Klein 
monopoles \cite{Dowker:1994up} is also good, as is the nucleation of spherical bubbles of 
Kaluza-Klein $(n-5)$-branes in magnetic fluxbrane backgrounds 
\cite{Dowker:1996sg}. The decay 
of the Kaluza-Klein vacuum \cite{Witten:1982gj}
is good, which is slightly disappointing: one 
might have hoped that it would be stabilized by these considerations. 
We know, however,
that the cobordism for KK vacuum decay is the same as that for 
pair production of KK monopoles \cite{Dowker:1995gb}
and so if the latter is a good process so 
must the former be. 

The Big Bang, or creation of an $(n-1)$-sphere from nothing via the 
yarmulke, is good. Notice that in this way of treating topology change as a 
sequence of elementary changes, the universe, if created from nothing, must
start off as a sphere. No other topology is cobordant to the empty 
set via an elementary cobordism. The conifold transition in string theory
\cite{Greene:1995hu} where 
a three-cycle shrinks down to a point and blows up again as a two-cycle is
good. Indeed, the shrinking and blowing up process traces out the 
seven-dimensional 
cobordism (each stage of the process is a level surface of a 
corresponding Morse function) 
and the fact that it is a three-cycle that degenerates and a  
two-cycle that blows up tells us that the index of the cobordism is 
three. 

Bad topology changes include spacetime wormholes where 
an $S^3$ baby universe is born by branching off a 
parent universe, the epitome of a trousers cobordism. 
Stephen founded the study of baby universes and 
spacetime wormholes \cite{Hawking:1988ae} within the Euclidean quantum 
gravity framework 
where our present considerations do not apply.  
However, if one takes the view that Euclidean solutions -- instantons --
are to be thought 
of as devices 
for calculating transition amplitudes which are
nevertheless defined as sums 
over real, causal, spacetimes, then the badness of
the trousers would be
counterevidence for the relevance of spacetime wormholes.   

Another bad topology change is the pair production or annihilation
of topological 
geons, particles made from non-trivial spatial topology. This deals
a serious blow to the hope that the processes of pair production and
annihilation of geons could restore to geons the spin-statistics 
correlation that they lack if their number is fixed \cite{Dowker:1998ei}.

In $1+1$ and $2+1$ spacetime dimensions, all topology changes except for the
yarmulkes and their time-reverses are bad ones. This raises the
question, is this not in conflict with string theory and the finiteness
of topology changing amplitudes in $2+1$ 
quantum gravity \cite{Witten:1989sx}? For the former, Lenny 
Susskind has argued that the  
the infinite burst of energy when a loop of string splits into two 
can be absorbed as a renormalisation of the string coupling constant 
\cite{Susskind:2002}. 
Martin Ro\v{c}ek takes an alternative view, 
that in first quantized string theory, 
choosing to integrate over causal metrics on the world sheet would be
analogous to choosing paths in the SOH for relativistic  
quantum mechanics that move only forward in time which would be 
inconsistent \cite{Rocek:2002}.
In the latter case, it seems that in the first order, frame-connection formalism 
suitable for $2+1$ gravity, topology change can occur even as a classical
process since the relationship between the frame and 
connection is an equation of motion and can hold even at points where the 
frame is degenerate \cite{Horowitz:1991qb}. In this case, it is no surprise that 
quantum amplitudes for topology changing processes in $2+1$ gravity are 
non-zero. In fact, it becomes something of a puzzle why we do not see 
topology change on macroscopic scales 
all the time if it is an allowed classical process. 
It would seem that the first order formalism and the metric formalism are 
genuinely different theories of gravity and distinguishing between
them might be an observational issue.   

\section{Progress on the Borde-Sorkin conjecture}

Having looked at some of the consequences of the conjectures, 
we can ask how plausible they are. There is 
fragmentary evidence for the conjecture that causal discontinuity 
leads to badly behaved quantum fields but causally continuous 
topology changes allow regular quantum field behaviour \cite{Sorkin:1989ea,
Louko:1997jw}.  
A key  investigation that needs to be done is of
quantum field theory on a four dimensional spacetime with an index 
two point, which is conjectured to be regular. 

On the other hand we are well on the way to proving the 
Borde-Sorkin 
conjecture that Morse spacetimes are causally
continuous if and only if they contain no index $1$ or $n-1$ points.
I will sketch the basic ideas involved in the progress made
to date.

\begin{figure}[thb]
\centerline{\epsfbox{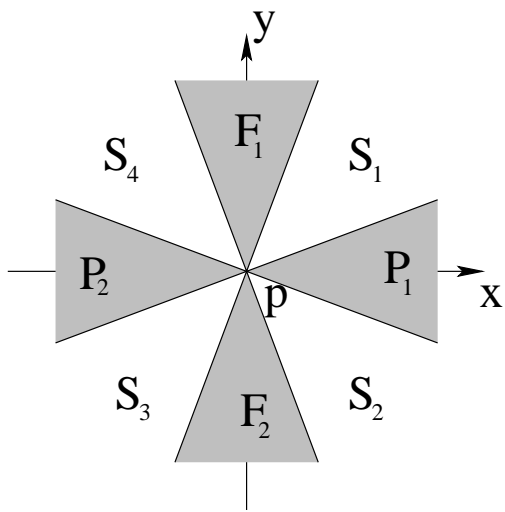}}
\caption{The causal structure in the neighbourhood of the 
 Morse point, $p$, of the $1+1$ trousers.
The past (future) of $p$
consists of the two regions 
$P_1$ and $P_2$ ($F_1$ and $F_2$). The ``elsewhere'' of $p$ divides into
four regions $S_1,\dots S_4$.
}
\label{fig:tpart}
\end{figure}

If we think about the causal structure around the Morse point, $p$, 
of the $1+1$ trousers, it seems intuitive that the causal past of 
$p$ should contain two separate parts, one down each
``leg'' of the trousers. And the causal future of the Morse point
also divides into two lobes, one up the front and one up the 
back of the trousers. It's also  intuitive that the causal 
discontinuities of the $1+1$ trousers  should be related to the  
disconnectedness of the causal past and future of $p$ in the 
neighbourhood of $p$. 
Flattening out the crotch region, we should obtain a causal structure that 
looks like that shown in figure \ref{fig:tpart}.  
There is a special metric for the $1+1$  trousers in which the causal 
structure  can be 
proved to be exactly as shown: the past (future) of the Morse point $p$
consists of the two regions
$P_1$ and $P_2$ ($F_1$ and $F_2$). 

There are two types of causal discontinuity here. The first type is 
when an observer starting in $P_1$, say, crosses the past light cone
of $p$ into $S_1$, say. As it does so 
the causal future of the observer,
which at first contains regions in both $F_1$ and $F_2$, jumps so that 
it no longer contains any points in $F_2$. 
The second type is when an observer in $S_1$, say, crosses the 
future light cone of $p$ into $F_1$. As this happens, the 
causal past of the observer which contained no points in $P_2$ 
suddenly grows to contain a whole new region in $P_2$. 

The special metric in which this behaviour can be demonstrated 
exactly generalizes to higher dimensions and all Morse indices. 
For dimension $n$ and index $\lambda \ne 0, n$ (no yarmulkes 
for now), the causal past and future of $p$ are obtained from
figure \ref{fig:tpart} by rotating it around the $x$-axis
by $SO(n-\lambda)$ and around the $y$-axis by $SO(\lambda)$.
We see that when $\lambda=1$ the past of $p$ remains in
two  disconnected   pieces and when $\lambda =n-1$ the 
future of $p$ remains in two pieces. But when $\lambda \ne 1, n-1$
then both the future and past of $p$ become connected sets. 

These suggestive pictures can be turned into a proof that, 
for these special metrics, index $1$ and $n-1$ Morse points 
produce causal discontinuities and the other 
indices do not. We can further show that, not just these 
special metrics, but any index $1$ and $n-1$ Morse 
metric is causally discontinuous. It is also true that any
Morse metric on the yarmulke is causally continuous. It 
remains to be proved that any Morse metric on 
a $\lambda \ne 0,1,n-1,n$ elementary cobordism is 
causally continuous. 

\section{Looking to the future}

I will end by speculating on possible implications that these results
might have for a particular proposal for quantum gravity, 
namely causal set theory, the approach 
championed by Rafael Sorkin \cite{Bombelli:1987aa, Sorkin:1990bh, Sorkin:1990bj}.
The basic hypothesis 
is that the underlying fundamental substructure of spacetime is 
a discrete object called a causal set, which is, roughly, a 
Planck density random sampling of
the causal structure of spacetime. Mathematically, 
a causal set, $C$, is a partial order 
(thus it satisfies the conditions of transitivity and irreflexivity  
given previously)  which is also ``locally finite'' meaning  
that the set $\{z\in C: x\prec z\prec y\}$ has finite 
cardinality for all pairs of points $x$ and $y$ in $C$.

How could a causal set hope to be the basic stuff of spacetime? How could 
such a truly discrete entity, with not a real number in sight, underlie a  
continuum spacetime with its topology, differential structure and metric?
The answer lies in work by Stephen and others that shows that for Lorentzian
manifolds satisfying a certain causality condition (marginally 
stronger than the absence of closed causal curves) the causal structure 
determines the metric up to an overall conformal factor. 
Since this result is not as well known as it should be, I'll 
give a few details here. Stephen proved that if
$(M,g)$ and $(M',g')$ are Lorentzian spacetimes and $f:M\rightarrow M'$ is
a homeomorphism where $f$ and $f^{-1}$ preserve future directed 
continuous null geodesics then $f$ is a smooth conformal 
isometry \cite{Hawking:1976fe}. Malament used this result to 
show that if $(M,g)$ and $(M',g')$ are past and future 
distinguishing spacetimes (this condition 
means that distinct points have non-equal 
pasts and futures) and $F:M\rightarrow M'$ is
a bijection such that $x \in I^+(y)$ if and only if
$F(x) \in I^+(F(y))$ $\forall x,y \in M$, then 
$F$ is a smooth conformal isometry \cite{Malament:1977}. 
Finally one can show that a bijection that preserves the 
causal structure, $J^+$ ({\it {i.e.}} $F:M\rightarrow M'$ 
such that $x \in J^+(y)$ if and only if
$F(x) \in J^+(F(y))$ $\forall x,y \in M$) also preserves the 
chronological structure $I^+$ when the spacetimes are 
distinguishing \cite{Levichev:1987}. 
Notice that in the final form of the result, the bijection is 
not required even to be continuous or have any properties apart from
being causal structure preserving. 

Given this powerful result, it seems plausible, even reasonable
to suppose 
that a Planck scale ``discretization'' of the causal structure
should encode all information 
about a spacetime at length scales
above the Planck length. The missing 
conformal factor, or volume information, is fixed by making the 
correspondence that the volume of a region counts the number of 
causal set elements contained in the region.   

The causal set hypothesis is both conservative and radical. It is
conservative in that it takes
the belief that many workers in quantum gravity hold
that spacetime is discrete at the fundamental level
and the theorem that causal structure is nearly all the metric and puts them
together in the most obvious way: the underlying
structure is a discretization of the causal structure.
The hypothesis then ties up, in a most satisfying way,
the question of how the remaining spacetime information
is provided, because the correspondence of Volume $\sim$ Number
can only be made  due to the discreteness of the
causal set. The details
of the hypothesis include the prototype of a 
solution to the knotty problem of how
to discretize spacetime
whilst maintaining local Lorentz invariance, which solution
-- roughly that the discretization is random -- is
conjectured by Sorkin to be essentially unique.
Causal set theory
is nevertheless radical because it proposes that, fundamentally,
there is only a
local finitude with a
partial order. Dimension, manifold, topology, differentiable structure,
spacetime metric, spacetime causal structure,
perhaps also matter, would all be unified in terms of 
order. To whet the appetite even further, let me 
mention that causal set theory 
was used to make a surprising prediction which 
has subsequently been verified by observation: namely a  
prediction of the current order of magnitude of a non-zero cosmological 
constant \cite{Sorkin:1990bh,Sorkin:1990bj,Sorkin:1997gi}. 

Could the picture of topology change I have sketched 
tell us anything about
causal set calculations or vice versa? To see how it could, I will 
discuss two examples.
 
\begin{figure}[thb]
\centerline{\epsfbox{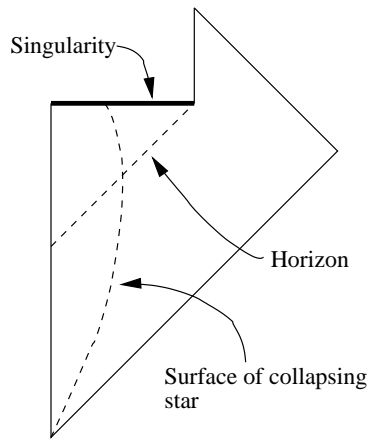}}
\caption{
The Carter-Penrose conformal diagram of the formation and evaporation of a black hole.
}
\label{fig:bh}
\end{figure}

One is the process of black hole evaporation which Stephen discovered, 
the understanding of 
which is going to play a key role in the development of any 
successful theory of quantum gravity. As far as the continuum theory goes, 
the nearest we can come to a spacetime description of black hole 
formation and evaporation is the conjectured 
Carter-Penrose diagram shown in figure 
\ref{fig:bh}. 
What could be the description in quantum gravity of this crucial 
process? One motivation for Stephen's study of baby universes 
was that a spacetime wormhole, or trousers, was the most obvious 
candidate for a spacetime description in the Euclidean approach, 
with the matter falling into and the radiation emitted from the 
black hole residing in the baby universe. 
The approach we have taken here, however, suggests that trousers 
cannot occur and so we must look for a different 
description. 

According to the causal set hypothesis, the underlying reality is
a causal set, we know that much. Where we have a continuum description, 
the causal set must be manifold-like and where the continuum 
description breaks down, close to the singularity, the causal set will
have no continuum approximation and only the causal set itself will 
be a good description of what is happening there. We don't yet have a 
quantum dynamics for causal sets but we do have a family 
of physically motivated ``warm up'' dynamical models to play with.  
These are classical, stochastic models in which a causal set grows,
element by element, in a probabilistic way that respects 
(the discrete versions of) general covariance and classical causality,
the ``Rideout-Sorkin'' (RS) models \cite{Rideout:1999ub}.
Joe Henson has proved that in these RS models every element in
a causal set almost surely has an element to its future, 
with the corollary that 
every point has an infinite future set \cite{Brightwell:2002}.
This result might be called, ``Causal Immortality''.

If this prediction persists in the quantum theory of causal sets we 
will then know what to expect in a causal set description of black hole
formation and evaporation. The causal set will have a continuum approximation
corresponding to almost all of the Carter-Penrose conformal diagram
but close to  
the singularity the causal set 
will cease to be manifold-like. It 
won't, however, end there: 
the causal set will 
extend on into the future. It's possible that 
the part of the causal set that 
is ``born'' in the singularity will continue to be non-manifold-like forever
but, alternatively, it may become manifold-like again.
(Note, these words ``continue'', ``forever'' and ``become'' are merely suggestive, they 
are not supposed to imply a physically meaningful background
``time''.)
In the latter case it would then be a ``baby universe'', disconnected  from the
region in which the black hole formed and evaporated, 
though not a baby universe created by a continuum wormhole. 

It is tempting to speculate yet further that this might realise 
something like Smolin's idea of cosmological natural selection \cite{Smolin:1992us}, 
with new universes with different coupling 
constants being created, not inside black holes but
when black holes evaporate. 
The causal structure of the black hole singularity in the
formation-evaporation spacetime would be the following. If 
the singularity represents a single causal set point $s$  
and if $F(s) = \{ x\in C: s\prec x\}$ is the future set 
of $s$, then any point $x\in F(s)$ would have the property that 
a point related to $x$ must be related to $s$. This says that nothing
can influence the future of $s$ except through $s$ and means that 
$s$ is a so-called ``past post'' as far as $F(s)$ is concerned
(a ``post'' is a point in a causal set to which all points in the 
set are related).  
Now, posts are responsible for a cosmological renormalisation 
of the coupling constants which characterize the dynamics of causal 
sets, at least in the classical RS models \cite{Sorkin:1998hi,Martin:2000js}
and it seems that a similar renormalisation would take place 
when there's a past post. 
 
The second example is an investigation into how causal discontinuity 
might be ruled out by causal set dynamics. First we need a characterization
of causal discontinuity in terms of the causal order alone. 
This might be done by looking at the various equivalent conditions  
for causal discontinuity in the continuum 
that Stephen and R.K. Sachs have provided 
\cite{Hawking:1974}.  Alternatively, thinking about 
causal sets which arise by randomly 
sampling causally discontinuous spacetimes, 
it seems that one possible characteristic that they share is the 
existence of an element $x$, which has a (past or future) relation, which 
if severed would make a big change in the size of the past or future 
set of $x$. One would have to make this more precise but intuitively 
it seems to make sense. Then one would hope to be able to predict that 
causal sets with such a property do not occur. This would have to 
be done in the, as yet unknown, quantum theory of causal sets but 
to see broadly how it might work we could  ask whether we are able 
make such a prediction in the RS models.  
In these models, one can prove things like Henson's result that 
every point has a future relation.  Graham Brightwell has also  
shown that an infinite antichain (an 
antichain is a totally unordered subset) almost surely doesn't occur
\cite{Brightwell:2002}. It seems possible that one could prove something like
``the probability that a point $x$ will have a future set which has
a large subset which will be cut off from $x$ by the severing of 
a single relation is zero.''

So, for the future, there is much to be done on the specific conjectures
mentioned, in order to put this whole approach to topology change
on a firmer footing. And for the broader picture,
I think that things look very interesting and promising for the
causal set approach to quantum gravity.
Although Stephen has not
worked directly on this area, his work on
global causal analysis is one of the crucial underpinnings
of the programme.
His work  is  central to the proof that the causal structure
of a spacetime encodes most of the information about that spacetime.
He has also been one of the main proponents of the Sum Over Histories
approach to quantum gravity which  will be at the heart of developments
on a quantum dynamics for causal sets. This is because 
a causal set has an essentially ``spacetime'' character: 
it's hard to see how one could make any headway with an attempt 
at the space+time split required for a canonical quantization for 
example. 
 In a sense, causal sets provide 
an explanation for Stephen's Chronology Protection Conjecture
because one can ``predict'' that there will be no CTC's (this is 
slightly more than a
case of simply assuming what we want to prove -- the irreflexivity condition
on the causal set can be dropped but a spacetime with CTC's will 
nevertheless 
never be an approximation to a transitive digraph even when 
it has closed loops \cite{Sorkin:1997}). 
Of course, Stephen's work on
black hole thermodynamics will have a major role to play in any
proposed theory of quantum gravity but it is particularly
important in causal set theory since one of the main
motivations for believing in an underlying  discrete structure at
all is the
finiteness of the black hole entropy \cite{Sorkin:1997gi}. 
Altogether, it wouldn't be going too far to say that the
influence of Stephen's work can be seen in the very foundations of
the causal set approach. I hope there will be some successes of causal set
theory to report for Stephen's 70th birthday.

\bibliography{top}       
\bibliographystyle{unsrt}

\end{document}